\documentclass[runningheads,a4paper]{llncs}
\usepackage[titletoc,title]{appendix}

\usepackage{amssymb}
\usepackage{graphicx}

\usepackage{caption}
\usepackage{subcaption}
\usepackage{subfig}

\usepackage{amsmath}
\usepackage{setspace}

\newcommand{\DD}{\, \displaystyle}

\newcommand{\qtil}{\, \widetilde{\theta}_c }
\newcommand{\ktil}{\, \widetilde{k}_c }
\newcommand{\qmed}{\, \hat{\theta}_c }

\newcommand{\thmax}{\,\theta_{\textnormal{max}}}

\newcommand{\fracc}[2]{\, \displaystyle \frac{ #1}{ #2}}

\newcommand{\ave}[1]{\, \overline{#1} }

\newcommand{\morabba}[1]{\,\begin{flushright}
 \Rectsteel \\
\end{flushright}}

\newcommand{\CC}[2]{\, \binom{#1}{#2} 
}

\newcommand{\all}[2]{\,\begin{align}
                   #1 
                    \label{#2}
                   \end{align}
}
\makeatletter
\newcommand{\vast}{\bBigg@{4}}
\newcommand{\Vast}{\bBigg@{5}}

\makeatother
\begin{document}

\mainmatter  

\title{ Measuring the Generalized Friendship Paradox
in Networks with Quality-dependent Connectivity}

\titlerunning{Measuring the Generalized Friendship Paradox}

%

\author{Naghmeh Momeni and  Michael G. Rabbat}
\authorrunning{Naghmeh Momeni and  Michael G. Rabbat}

\institute{Department of Electrical and Computer Engineering\\
McGill University, Montr\'eal,  Canada\\
\email{naghmeh.momenitaramsari@mail.mcgill.ca}, 
\email{michael.rabbat@mcgill.ca}}

%
%
%
\maketitle

\begin{abstract}
 The friendship paradox  is a sociological  phenomenon stating that most people have fewer friends than their friends do. The generalized friendship paradox refers to the same observation for attributes other than degree, and it has been observed in Twitter and scientific collaboration networks. This paper takes an analytical approach to model this phenomenon. We consider a preferential attachment-like network growth mechanism governed by both node degrees and `qualities'. We introduce measures to quantify paradoxes, and contrast the results obtained in our model to those obtained for an uncorrelated network, where the degrees and qualities of adjacent nodes are uncorrelated. We shed light on the effect  of the distribution of node qualities on the friendship paradox. We consider both  the mean and the median to  measure paradoxes, and compare the results obtained by using these two statistics. 
\end{abstract}
\onehalfspacing
\vspace{-7mm}
\section{Introduction}\label{intro}
 \vspace{-2mm}
The friendship paradox,  introduced by Feld~\cite{Feld}, is a sociological observation that says most people are less popular than their friends on average.  It is called a `paradox'  because, while  most people believe that they are more popular than their friends~\cite{Zuck},  Feld observed that the converse is actually true.
 There are more recent observations agreeing with Felds', that study online environments.
For example on Twitter,  people you follow and also your followers have, on average, more followers than you do. They also follow more people than you do~\cite{Hodas_twitter}. On Facebook, your friends have, on average, more friends than you do~\cite{Ugander}.

The friendship paradox is about the inter-nodal inequality of the degrees. What happens if we consider other attributes? This is the focus of the `Generalized Friendship Paradox'~\cite{GFP_nature,GFP_PRE}. For example on Twitter,  your friends on average tweet more and also share more viral content than you~\cite{Hodas_twitter,Kooti_weird}. In the scientific collaboration networks your collaborators have on average more publications, more citations and more collaborators than you do~\cite{GFP_nature}.

The friendship paradox  has applications in spotting influential nodes. In~\cite{vaccine}, it is used for finding high-degree nodes for efficient vaccination. In order to sample a node with above average degree, a node is chosen uniformly at random and one of their neighbours will be sampled. In~\cite{GFP_Fowler},  the friendship paradox  is  used for the early detection of flu outbreaks among college students. In~\cite{GFP_disaster}, it is utilized to derive early-warning sensors during catastrophic events such as hurricanes. 

In this paper, first we explain a quality-dependent preferential attachment scheme introduced in~\cite{Other_us}. Then, we introduce measures to quantify the mean and the  median paradoxes.  In Section 4 these measures are computed numerically on the networks generated with the quality-dependent  model and also uncorrelated networks. We compare the results obtained in these networks using both the mean and the median statistics. Furthermore, we study the effect of node quality distribution on the quality and friendship paradoxes.
\vspace{-6mm}
\section{Model, Notation and Terminology}
\vspace{-4mm}
We consider a \emph{quality-based preferential attachment} (QPA) model, identical   to the  model proposed and analysed in~\cite{Other_us}. It is similar to the Barabasi-Albert model~\cite{BA_1}, but incorporates node qualities. Each incoming node has $\beta$ links, and a discrete quality $\theta$ drawn from a distribution $\rho(\theta)$ that is assigned to it upon birth. The probability of an existing node $x$ with degree $k_x$ and quality $\theta_x$ (at the instant) receiving a new link is proportional to $k_x+\theta_x$.


Once assigned, the quality of a node does not change. We denote the mean of the  quality distribution  by $\mu$.  Following~\cite{Other_us}, as the number of nodes tends to infinity,  $P(k,\theta)$, the fraction of nodes with degree $k$ and quality $\theta$  is given by:
\all{
P(k,\theta)
 = 
 \rho(\theta) \big( 2+\frac{\mu}{\beta}\big) 
  \frac{\Gamma(k+\theta)}{\Gamma(\beta+\theta)}
 \frac{\Gamma \big(\beta+\theta+2+\fracc{\mu}{\beta} \big) }
{\Gamma \big( k+\theta+3+\fracc{\mu}{\beta} \big) }  
  u(k-\beta)
.}{Pkth_fin}
 In~\cite{Other_us} the nearest-neighbor distribution, i.e., the fraction of neighbors of a node with degree $k$ and quality $\theta$ who has degree $\ell$ and quality $\phi$  is given by: 
\all{
  &P(\ell,\phi|k,\theta)=
\fracc{\rho(\phi)  }{k}
  \frac{ \Gamma \left(k+\theta+3+\fracc{\mu}{\beta} \right)}{\Gamma\left( k+\theta+3+\frac{\mu}{\beta}+\ell+\phi  \right)  }
\frac{   (\ell-1+\phi)! }{  (\beta-1+\phi)!  }  
\Gamma \left(\beta+2+\phi+\fracc{\mu}{\beta} \right)\times
\nonumber \\ 
&
\resizebox {0.93\linewidth}{!}{$
\left[ \DD
 \sum_{j=\beta+1}^{k} \frac{\Gamma \left(j+\theta+2+\fracc{\mu}{\beta}+\beta+\phi \right) \DD \CC{k-j+\ell-\beta}{\ell-\beta}
}{\Gamma \left( j+\theta+2+\fracc{\mu}{\beta} \right)\Gamma \left(\beta+2+\phi+\fracc{\mu}{\beta} \right) } 
 +  \DD \sum_{j=\beta+1}^{\ell} \frac{\Gamma \left(j+\theta+2+\fracc{\mu}{\beta}+\beta+\phi \right)\DD \CC{\ell-j+k-\beta}{k-\beta}
}{ \Gamma \left( j+\phi+2+\fracc{\mu}{\beta} \right)\Gamma \left(\beta+2+\theta+\fracc{\mu}{\beta} \right) } 
\right].$}}{NNFDD}

\section{Measures of Friendship and  Quality Paradoxes }
\vspace{-2mm}
By marginalizing the joint distribution $P(k,\theta)$ we can find the degree distribution, denoted by $P(k)$.  Also, from the nearest-neighbor distribution~\eqref{NNFDD}, we can find the expected value of the qualities of neighbors of a node with quality $\theta$ and also the expected value of the degrees of neighbors of a node with degree $k$. This allows us to investigate when the quality paradox (hereinafter QP) and the friendship paradox (hereinafter FP) are in force, and which nodes in the network exhibit the paradox. 

Let us also define the `median' version of the paradoxes, following~\cite{Kooti_weird}. In the median  version, instead of the average values of quality or degree of neighbors, we use the median values. A node experiences the median QP (FP), if its quality (degree) is less than the quality (degree) of at least half of its neighbors.   

Throughout the paper,    the superscript NN denotes  Nearest-Neighbor. 
 Let us denote the median  operator  by $M \{\cdot\}$.  For example, $M\{\phi^{NN} | \theta\}$ denotes the median value of $\phi$ under the distribution $P(\phi | \theta)$, and is a function of $\theta$.  Also note that every measure we introduce here is by nature a function of the parameters of the quality distribution. For example, if the exponential decay quality distribution is considered, the measures will depend on the decay factor. We denote the parameter of the quality distribution by $x$. Using  this  notation, we define the critical values for the mean and the median paradoxes as follows: 
\all{
\resizebox{.92\linewidth}{!}{$
\textnormal{mean:}
\begin{cases}
\widetilde{\theta}_c (x)\stackrel{\text{def}}{=} \max \bigg\{  \theta \Big| \theta <E \{ \phi^{\textnormal{NN}}|\theta \} \bigg\}
 \\
\widetilde{k}_c (x)\stackrel{\text{def}}{=} \max \bigg\{  k \Big| k <E \{ \ell^{\textnormal{NN}}|k \} \bigg\}
\end{cases}
,
\textnormal{median:}
\begin{cases}
\hat{\theta}_c  (x)\stackrel{\text{def}}{=} \max \bigg\{  \theta \Big| \theta < M \{ \phi^{\textnormal{NN}}|\theta \} \bigg\}
\\  
\hat{k}_c  (x)\stackrel{\text{def}}{=} \max \bigg\{  k \Big| k < M \{ \ell^{\textnormal{NN}}|k \} \bigg\}
\end{cases}$}
.}{measures}
In other words, $\widetilde{\theta}_c (x)$ is the highest quality that a node can have, given that its quality is lower than the average quality of its neighbors. Similarly, $\widetilde{k}_c (x)$ is the highest  degree  that a node can have, given that it exhibits the mean FP. For the median version of the paradox, we have $\hat{\theta}_c  (x)$ and $\hat{k}_c  (x)$. So  $\hat{\theta}_c  (x)$ is the highest quality that a node exhibiting the  median QP can have. 
Let us also emphasize that we use  the  following convention with regards to the median throughout the paper: the median of the probability distribution $g(x)$ (with CDF $G(x)$)  is the minimum value of $x$ for which $G(x) \geq \frac{1}{2}$. For example, for ${g(x)=\frac{1}{2} \delta[x]+\frac{1}{2}\delta[x-5]}$, the median is $x=0$. 

We now define similar quantities  for an `uncorrelated network'. In this network the qualities are assigned to nodes in an identical way to the QPA model, but the attachment of new nodes to existing nodes depends on neither the degrees nor the qualities of the existing nodes. In this network the properties of a node are uncorrelated with the properties of its neighbors. We denote this case by superscript $u$. For this network we have  ${P^u(\ell,\phi|k,\theta)=P(\ell,\phi)}$ and ${P^u(\phi|\theta)=\rho(\phi)}$. For the critical values of the mean and the median paradoxes, we have:
\all{\resizebox{0.93\linewidth}{!}{$
\begin{cases}
\widetilde{\theta}_c^u (x) \stackrel{\text{def}}{=} \max \bigg\{  \theta \Big| \theta <E \{ \phi^{\textnormal{NN}}|\theta \} \bigg\}
=  \max \bigg\{  \theta \Big| \theta < \underbrace{E  \{ \phi  \}}_{=\mu} \bigg\} = \mu(x)-1
\\  
\hat{\theta}_c^u  (x)\stackrel{\text{def}}{=} \max \bigg\{  \theta \Big| \theta < M \{ \phi^{\textnormal{NN}}|\theta \} \bigg\}
= \max \bigg\{  \theta \Big| \theta < \underbrace{ M \{ \phi  \} }_{\DD=  \hat{\theta}}\bigg\}= \hat{\theta}(x)-1
\end{cases}
.$}}{measures_u}
Similarly, for degrees we have: $\widetilde{k}_c^u(x)= \ave{k}(x) -1$ and $\hat{k}_c^u  (x)= \hat{k}(x)-1$.

We are also interested in the fraction of all nodes that experience each type of paradoxes.  This is equal to the fraction of nodes with their attribute below the corresponding critical value.  We denote these quantities by:
\all{
\textnormal{mean:}
 \begin{cases}
 \widetilde{F}_{\theta}(x)= \DD \sum_{\theta \leq \bar{\theta}_c(x)} \rho(\theta) \\  
 \widetilde{F}_k(x) = \DD \sum_{k \leq \bar{k}_c(x)} P(k) 
 \end{cases}
 ,~~~~~~
\textnormal{median:}
\begin{cases}
\hat{F}_{\theta}(x) = \DD \sum_{\theta \leq \hat{\theta}_c(x)} \rho(\theta) \\  
\hat{F}_k(x)= \DD \sum_{k \leq \hat{k}_c(x)} P(k)
\end{cases}
.}{Fs_def}

\vspace{-6mm}
\section{Results and Discussion}
\begin{figure}[!Ht]
        \centering
      \begin{subfigure}[b]{.493  \textwidth}
              \includegraphics[width=  \textwidth,height=4cm]{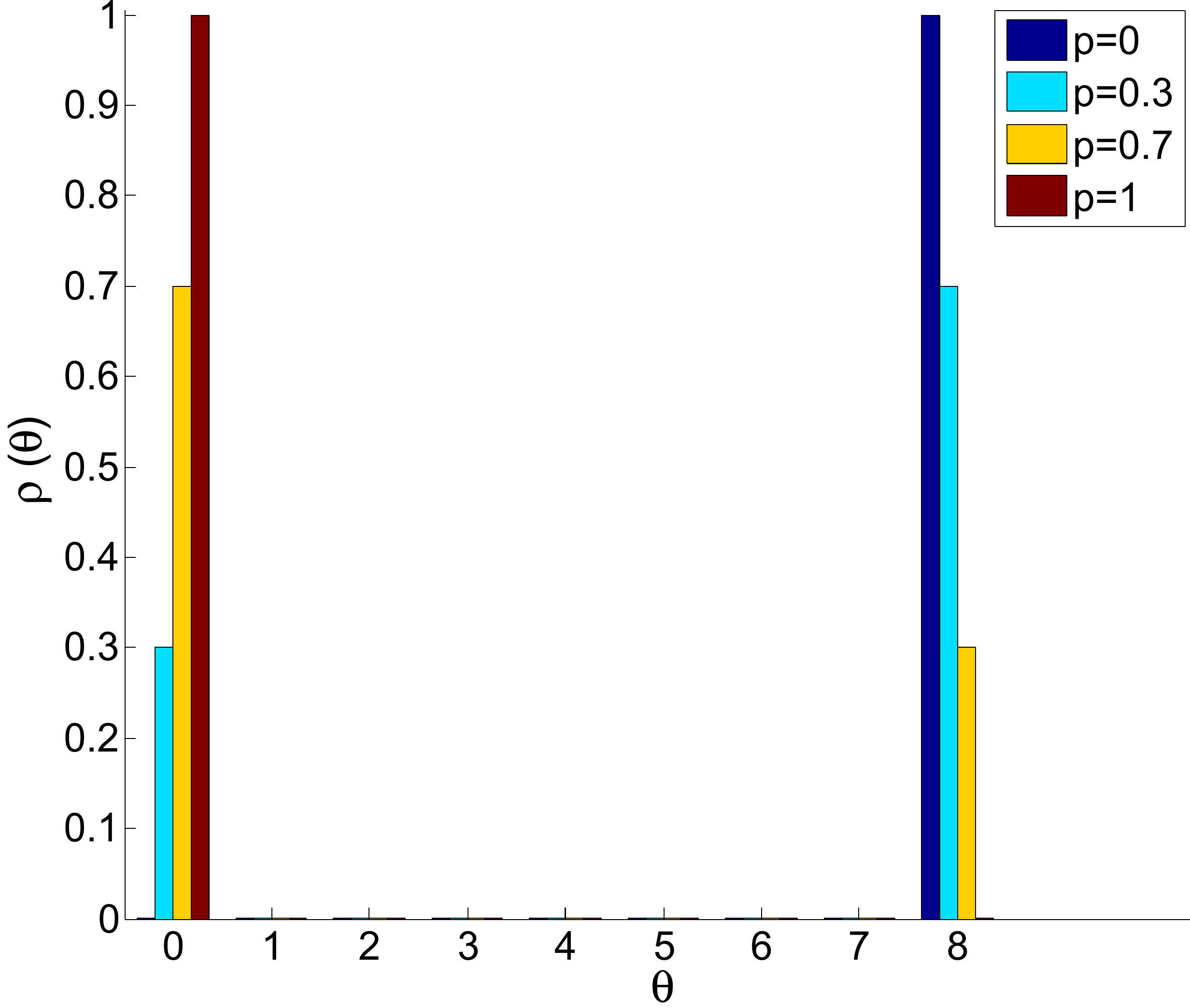}
                \caption{
  Bernoulli distribution with  ${p=0,0.3,0.7,0.1}$. The cases of ${p=0}$ and ${p=1}$ correspond to conventional Barabasi-Albert and shifted-linear preferential attachment networks, respectively. 
}
                \label{distribution_extreme}
        \end{subfigure}%
     ~~~   \begin{subfigure}[b]{0.493\textwidth}
                \includegraphics[width=  \textwidth,height=4cm]{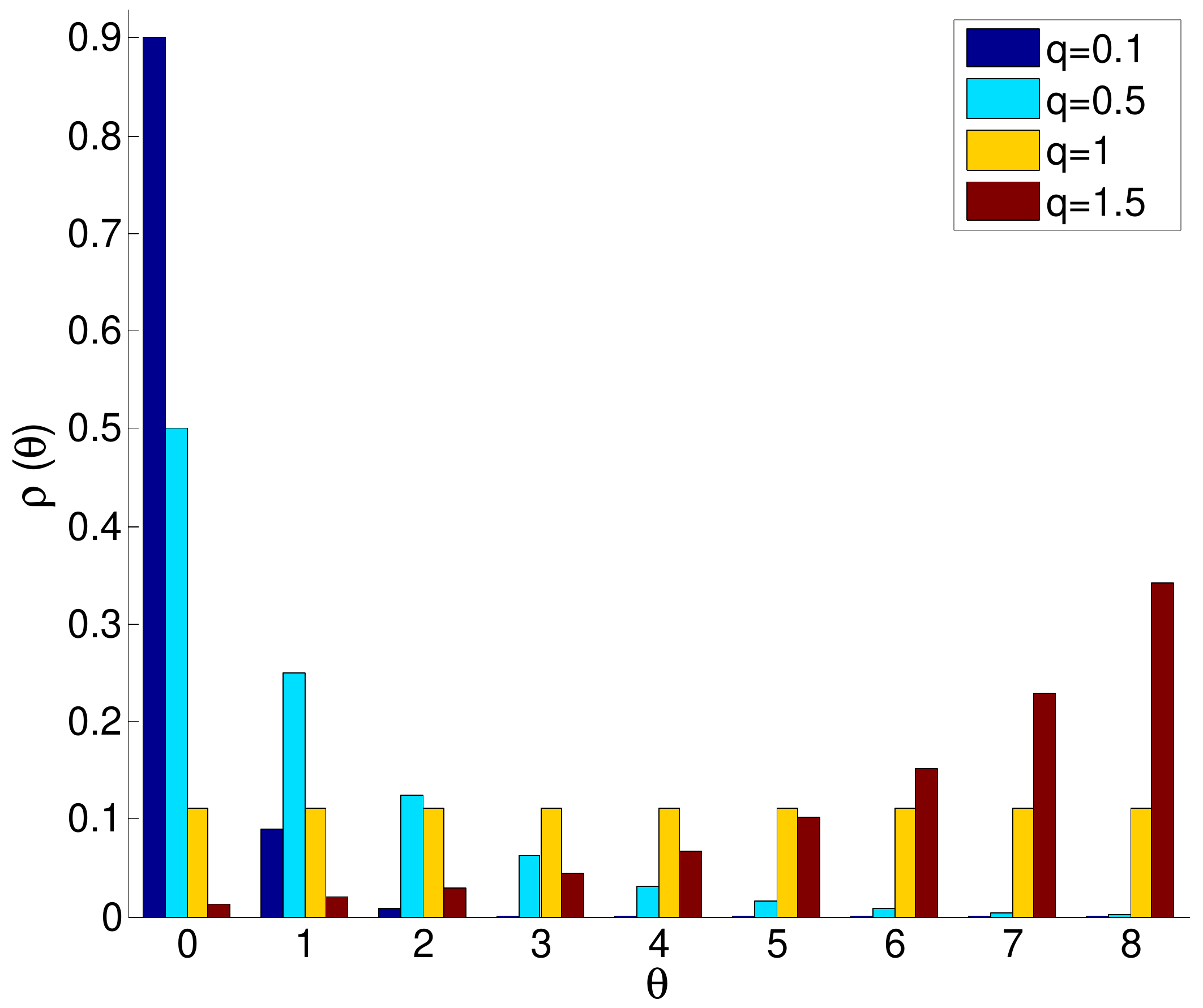}
                \caption{
Exponential distribution for decay factor ${q=0.1,0.5,1,1.5}$. The special case of ${q=1}$ corresponds to a uniform distribution supported in the interval ${0\leq \theta \leq \thmax}$.
}
                \label{distribution_decay}
        \end{subfigure}%
\caption{Examples of the quality distributions used in this paper with ${\thmax=8}$. Four instances of each type is depicted.}
\label{distributions}
\end{figure}

\vspace{-2mm}
In this paper we  consider two quality distributions for expository purposes. The first one is the Bernoulli distribution, where nodes have quality 0 (with probability $p$) or quality $\thmax$ (with probability ${1-p}$).  The other one is the discrete exponential distribution, with decay factor $q$. The probability of quality $\theta$ is proportional to $q^{\theta}$, and the maximum value of $\theta$ is denoted by $\thmax$.  Figure~\ref{distributions} depicts these quality distributions for four example values of $p$ and $q$.  Note that for $q<1$,  the exponential distribution is a decreasing function of quality  and ${\mu>\hat{\theta}}$, and for $q>1$, the distribution is increasing function of quality and ${\mu<\hat{\theta}}$. Also for the Bernoulli distribution note that, with the convention we use for the median, the value of the median is zero if $p \geq\frac{1}{2}$, and the median is equal to $\thmax$ if $p<\frac{1}{2}$. For each  distribution, we have numerically computed all the introduced measures for four different values of $\beta$ and four different values of $\thmax$. 

\begin{figure}
        \centering
      \begin{subfigure}[b]{.5 \textwidth}
              \includegraphics[width=  \textwidth,height=4.2cm]{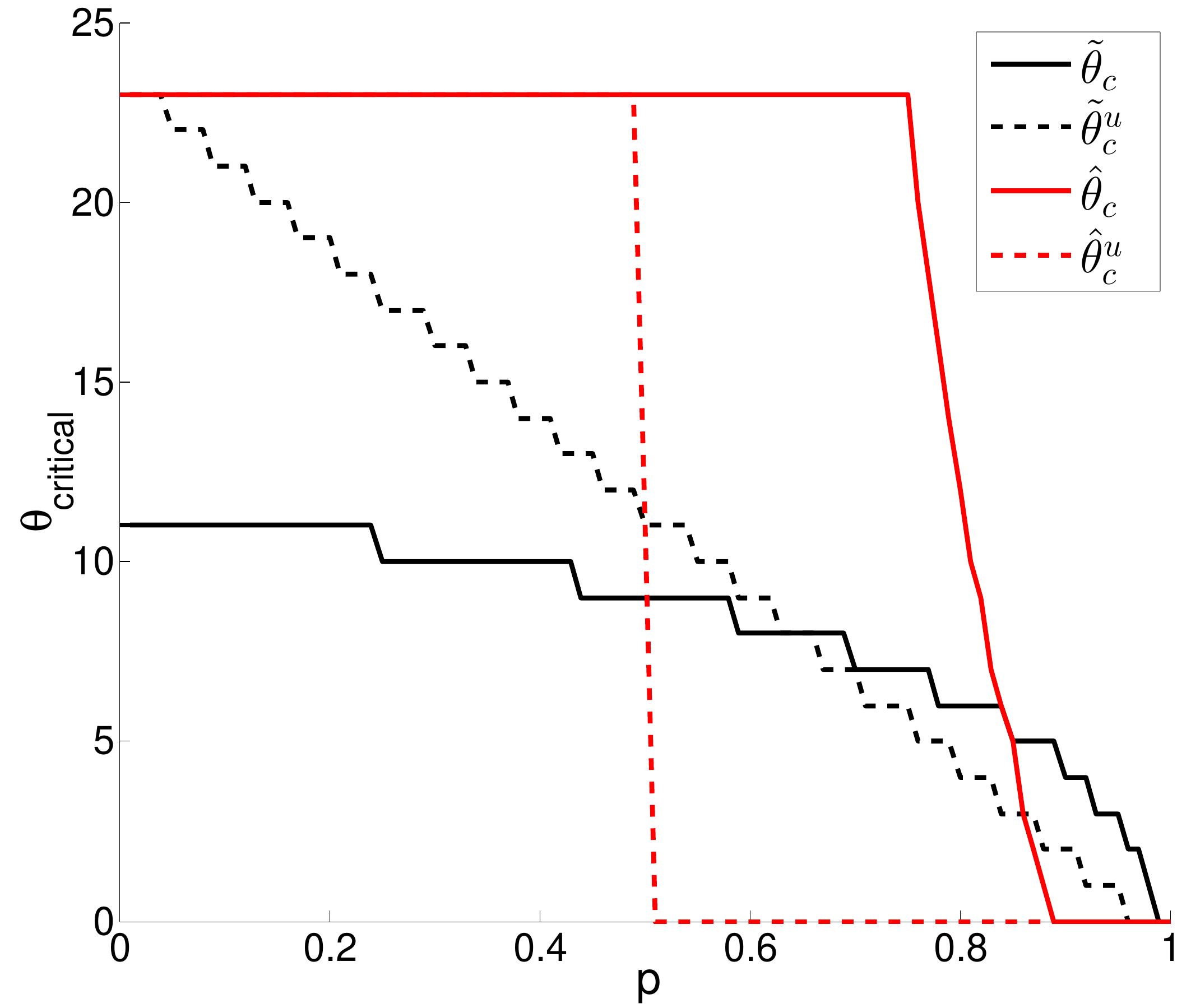}
                \caption{
Bernoulli, $\beta=2$ $\theta_{max}=24$
}
                \label{crit_ext_9}
        \end{subfigure}%
        \begin{subfigure}[b]{0.5\textwidth}
                \includegraphics[width=  \textwidth,height=4.2cm]{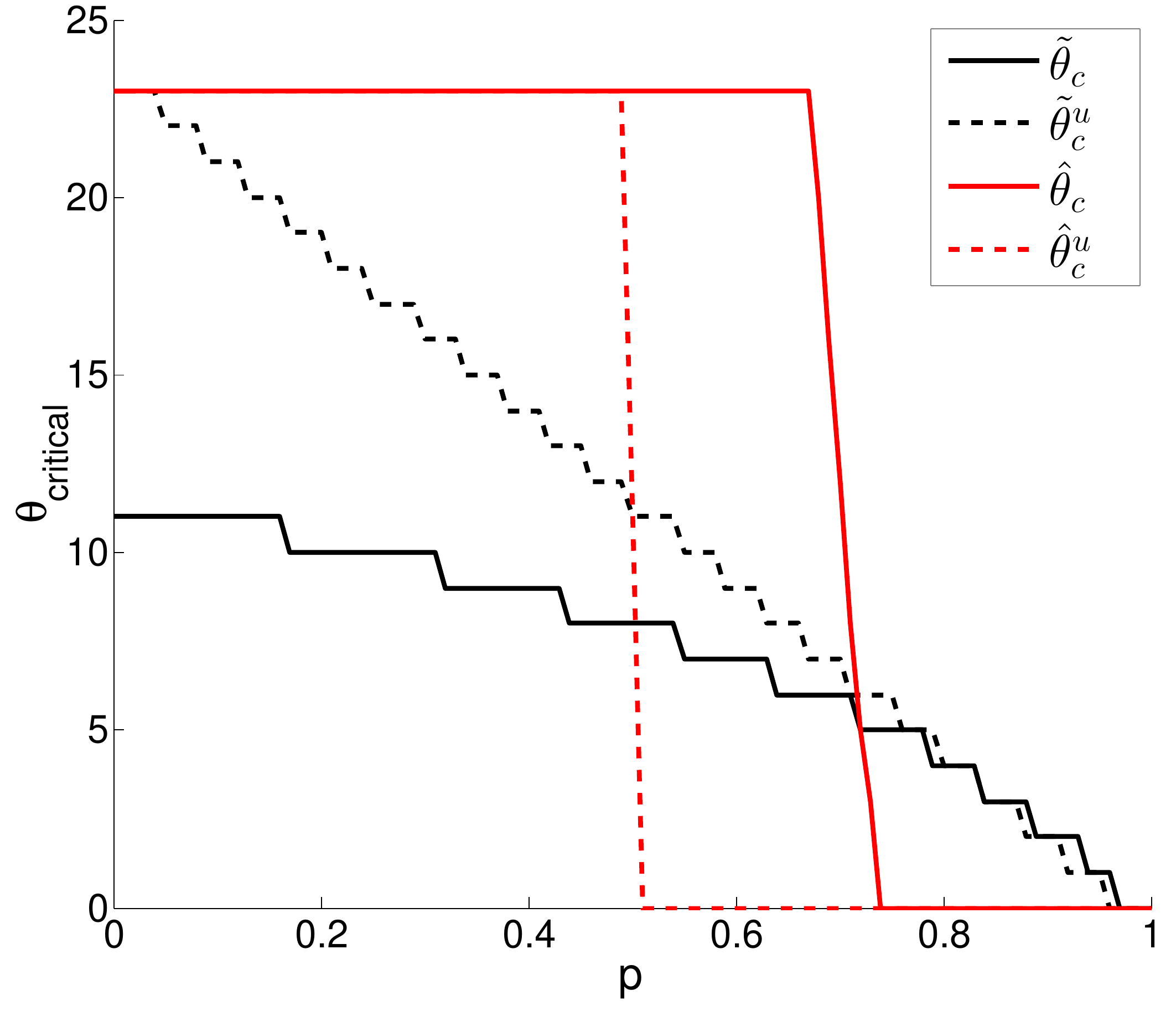}
                \caption{Bernoulli, $\beta=8$ $\theta_{max}=24$}
                \label{crit_ext_12}
        \end{subfigure}%
\\
 \begin{subfigure}[b]{.5 \textwidth}
              \includegraphics[width=  \textwidth,height=4.2cm]{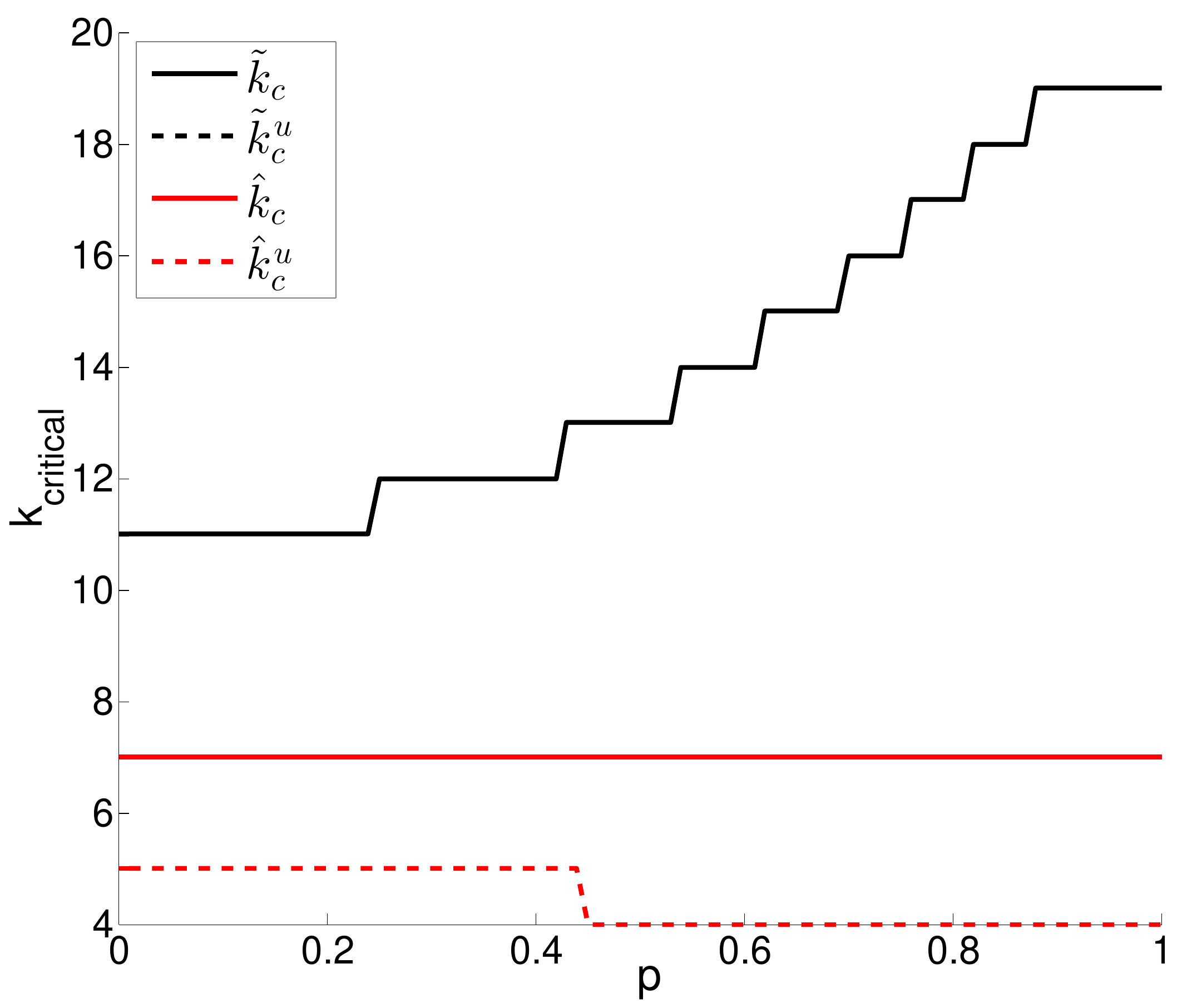}
                \caption{
Bernoulli, $\beta=4$ $\theta_{max}=16$
}
                \label{crit_ext_22}
        \end{subfigure}%
        \begin{subfigure}[b]{0.5\textwidth}
                \includegraphics[width=  \textwidth,height=4.2cm]{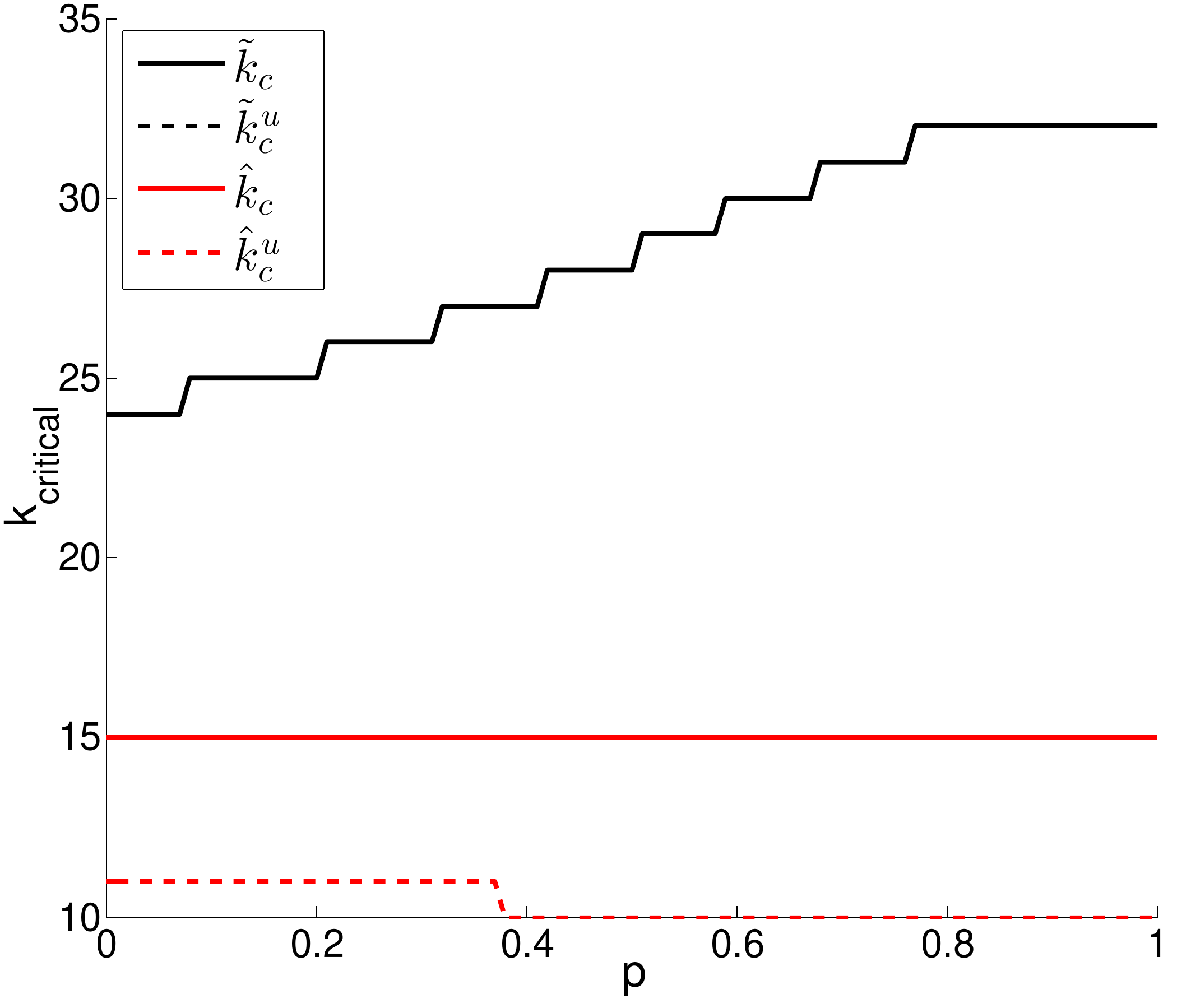}
                \caption{Bernoulli, $\beta=8$ $\theta_{max}=16$}
                \label{crit_ext_24}
        \end{subfigure}%
\\
 \begin{subfigure}[b]{.5 \textwidth}
              \includegraphics[width=  \textwidth,height=4.2cm]{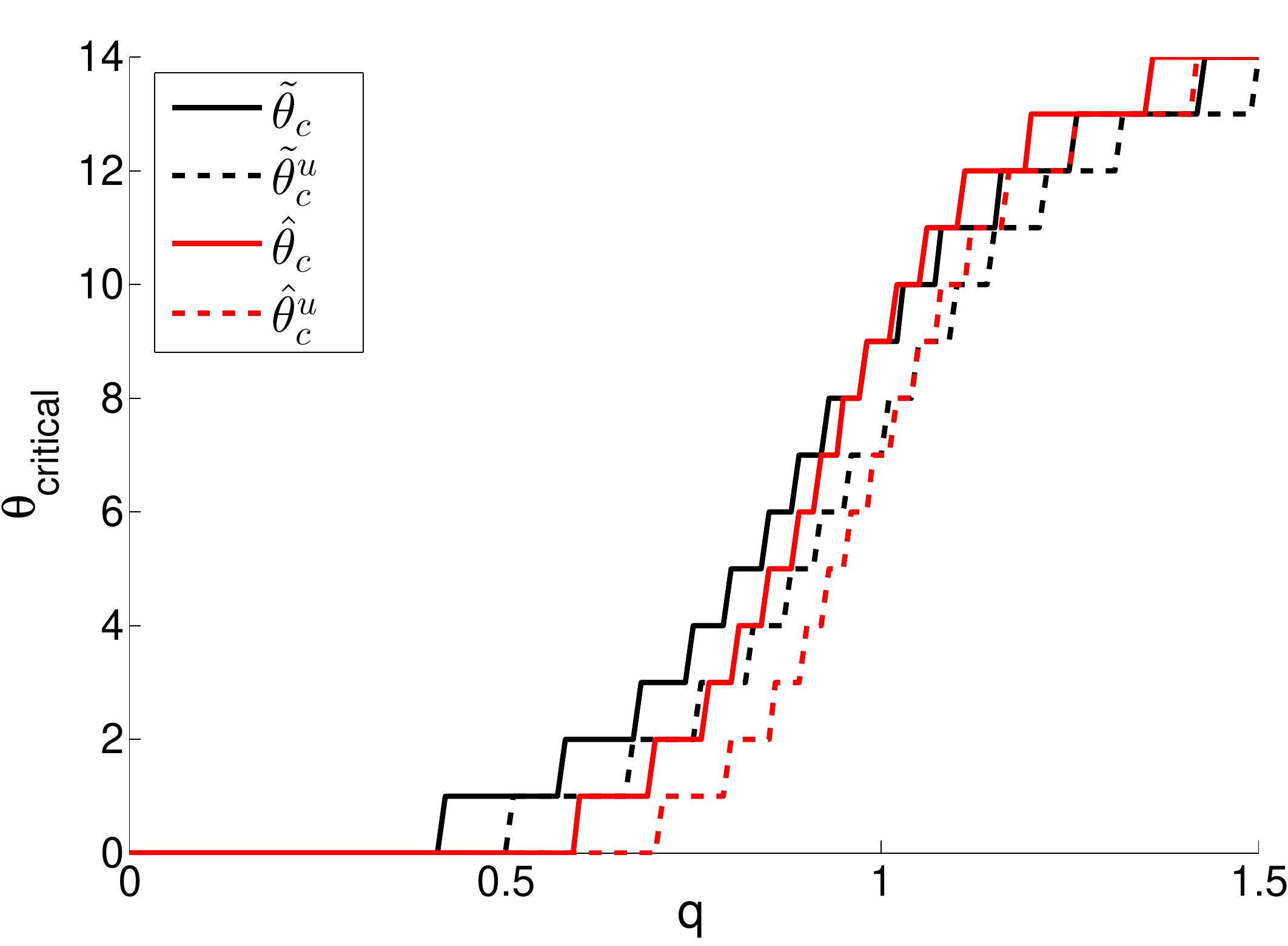}
                \caption{
Exponential, $\beta=2$ $\theta_{max}=16$
}
                \label{crit_decay_5}
        \end{subfigure}%
        \begin{subfigure}[b]{0.5\textwidth}
                \includegraphics[width=  \textwidth,height=4.2cm]{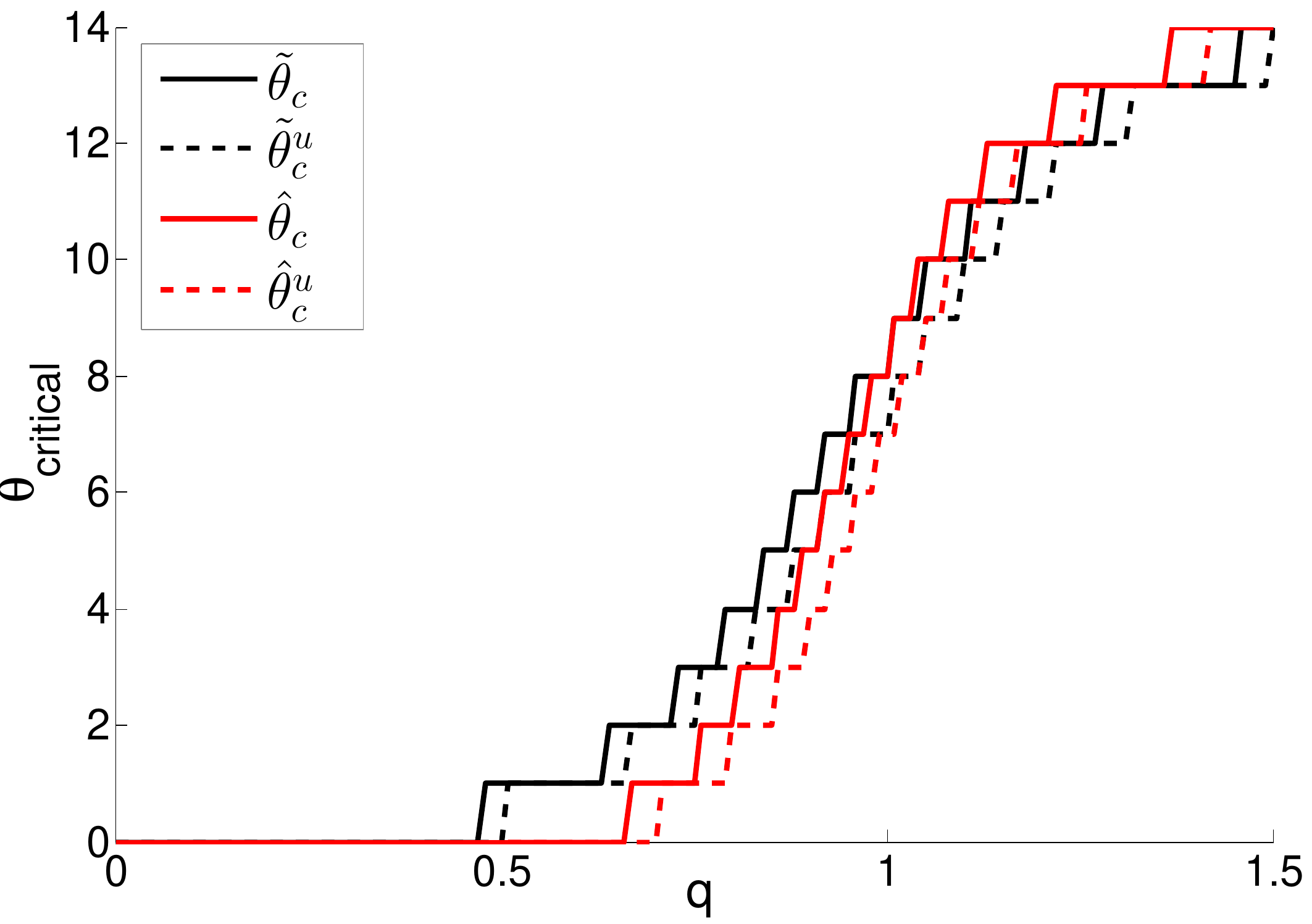}
                \caption{Exponential, $\beta=8$ $\theta_{max}=16$}
                \label{crit_decay_8}
        \end{subfigure}%
\\
 \begin{subfigure}[b]{0.5\textwidth}
                \includegraphics[width=   \textwidth,height=4.2cm]{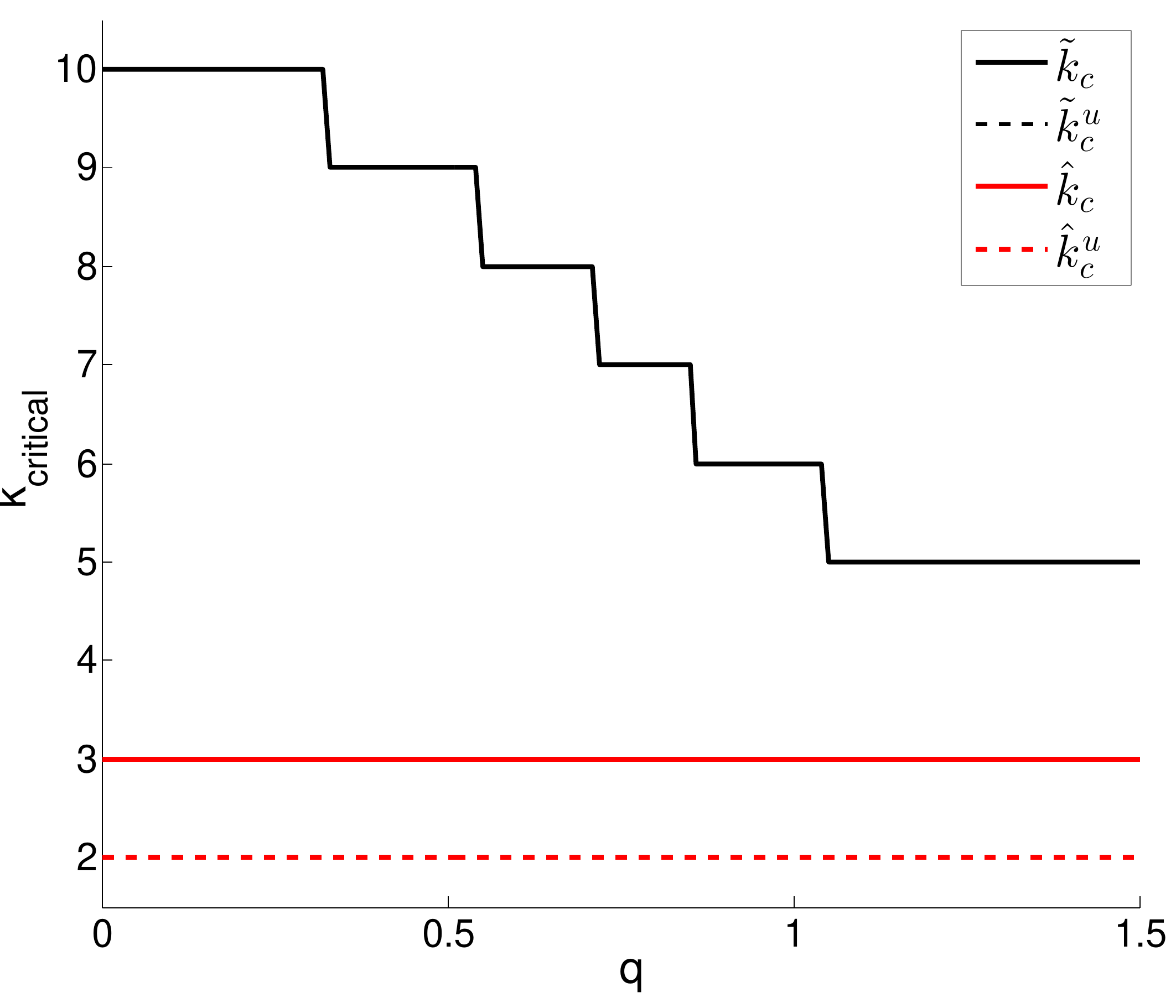}
                \caption{Exponential, $\beta=2$ $\theta_{max}=16$}
                \label{crit_decay_21}
        \end{subfigure}%
        \begin{subfigure}[b]{0.5\textwidth}
                \includegraphics[width=   \textwidth,height=4.2cm]{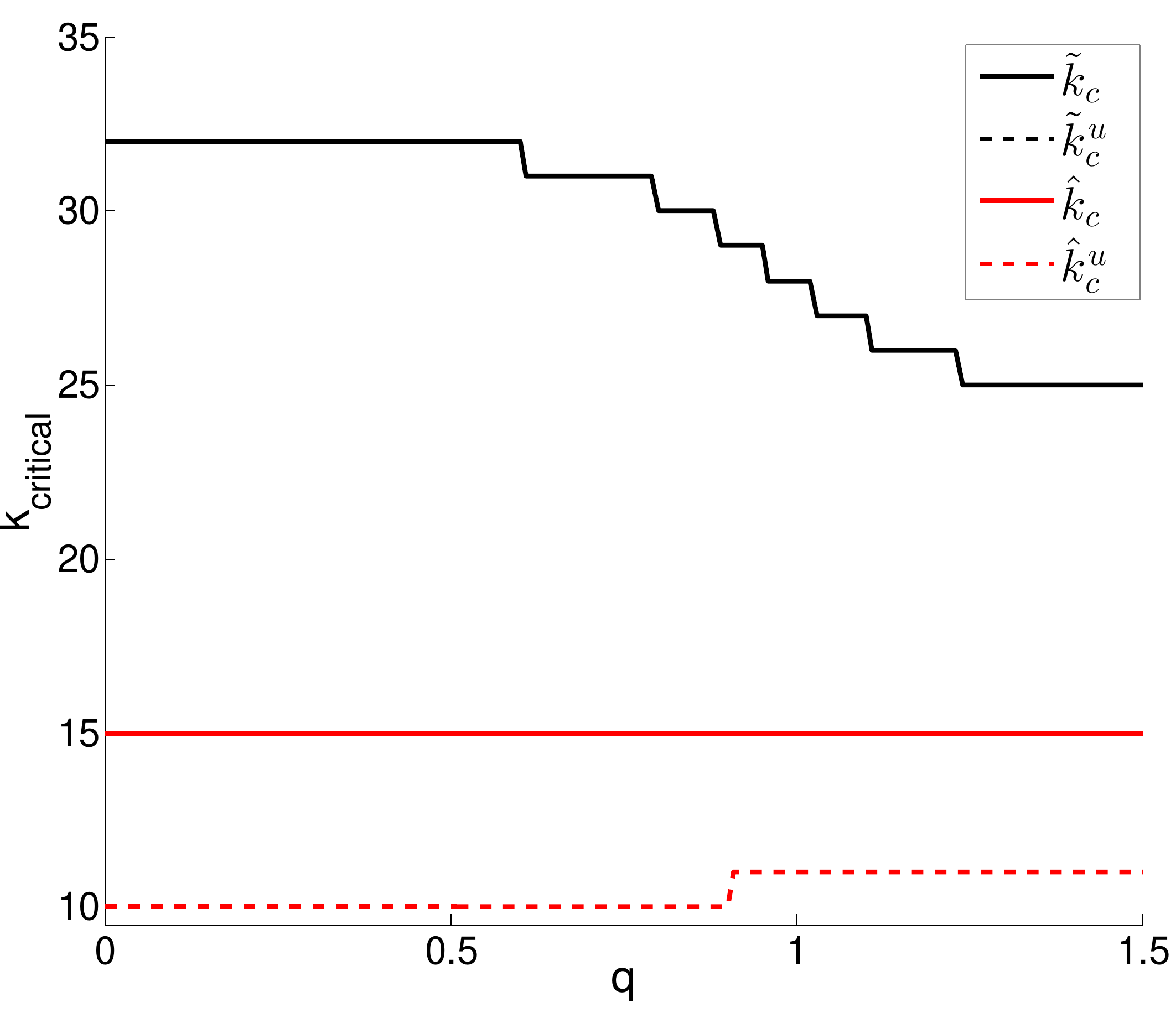}
                \caption{Exponential, $\beta=8$ $\theta_{max}=16$}
                \label{crit_decay_24}
        \end{subfigure}
         \caption{
      Critical values for quality and degree as defined in \eqref{measures} and \eqref{measures_u} computed for Bernoulli and exponential quality distributions.
          }
\label{criticals}
\end{figure}
Critical values obtained for two distributions are presented in Figure~\ref{criticals}. These values are computed using the closed form expressions mentioned in Section 2. From Figure~\ref{crit_ext_9} we  can learn about the differences between the networks that the QPA model generates and an uncorrelated  network. In an uncorrelated network  the probabilities of a random node being connected to a neighbor with quality $0$ and $\thmax$ are equal to  $p$ and ${1-p}$, respectively (regardless of the quality of the node). If the majority of the neighbors have quality zero ($p\geq 0.5$), the median is zero. Similarly, if the majority have quality $\thmax$  ($p< 0.5$), the median is $\thmax$. So if $p < 0.5$, nodes with qualities up to ${\thmax-1}$ experience the median QP and ${\qmed^u=\thmax-1}$. Conversely, if $p \geq 0.5$,  ${\qmed^u=0}$. This explains the abrupt drop in $\qmed^u$ in Figure~\ref{crit_ext_9}. On the other hand, in the QPA model, this transition takes place at a $p$ greater than $0.5$. This means that upto some point beyond $p=0.5$, although the probability of $\theta=0$ is higher than that of $\theta=\thmax$, the majority of the friends of each node have quality $\thmax$. There is a region for  $p>0.5$, where the majority of the network have quality zero, but the majority of the neighbors of most nodes have quality $\theta_{\max}$. This indicates quality disassortativity, since low quality nodes are mostly connected to nodes with high qualities.

For the mean version of the QP, we consider the  example  case of $p=0.2$ for discussion. In an uncorrelated network, each node (with any quality) is connected to neighbors with quality $0$ and $\thmax$ with probabilities $0.2$ and $0.8$, respectively. So the average of the qualities of its neighbors is $0.8 \thmax $. So nodes with quality less than ${ 0.8 \thmax }$ experience the mean QP.
  On the other hand, in the QPA model $\qtil<\qtil^u$ at   $p=0.2$. This means that  nodes whose qualities are between $\qtil$ and $\qtil^u$, do not experience the mean QP in the QPA model (while they do experience this paradox in the uncorrelated case). We deduce that these nodes are connected to quality zero nodes with a higher probability than $0.2$. This reduces the average quality of their niehgbors. Now consider the example case of $p=0.8$. In this case, $\qtil^u<\qtil$. This means that nodes with quality between $\qtil$ and $\qtil^u$ experience the mean QP in the proposed model, while they do not experience it in the uncorrelated case. In an uncorrelated network these nodes would be connected to zero and $\thmax$ quality nodes probabilities $0.8$ and $0.2$, respectively. However, in the QPA model, these nodes are connected to nodes with quality $\thmax$ with a probability higher than $0.2$, and this increases the average quality of their neighbors, making them subject to the mean QP. 

Comparing Figure~\ref{crit_ext_12} with~\ref{crit_ext_9} we observe the curves are similar, but the difference between the QPA model and the uncorrelated case is smaller in Figure~\ref{crit_ext_12}. For example, the drop in the $\qmed$ curve is closer to the drop in $\qmed^u$ for the uncorrelated case. We conclude that increasing $\beta$ decreases the difference between the QPA model and the uncorrelated  case. 

In Figure~\ref{crit_ext_22}, critical degrees  are depicted. It can be observed that as $p$ increases, $\ktil$ increases.  Comparing Figures~\ref{crit_ext_22} and~\ref{crit_ext_24}, we observe that all the critical degrees are greater in the case of $\beta=8$ than $\beta=4$. Also the range of node degrees experiencing any type of paradoxes is wider in the $\beta=8$ case.

From Figure~\ref{crit_decay_5}, we observe that for fixed decay factor, $\widetilde{\theta}_c \geq \widetilde{\theta}_c^u$ and  $\hat{\theta}_c \geq \hat{\theta}_c^u$. This means that there exist  values of $\theta$ that in the uncorrelated network  experience QP, but in the proposed model they do not. So the range of possible values of quality that experience the QP is wider in the QPA model than in uncorrelated networks. This argument holds for both mean  and median paradoxes. 

We also observe from Figure~\ref{crit_decay_5} that for $q<1$, $\qtil \geq \qmed$ and $\qtil^u \geq \qmed^u$. Both of these inequalities flip in the case of $q>1$. The main cause of this change of regime is the difference between the shape of the quality distribution for $q>1$ and $q<1$. When $q<1$, the median paradox is stronger (using the terminology of~\cite{Kooti_weird}), that is, the median paradox applies to a smaller range of qualities than the mean paradox (for both the uncorrelated network and the QPA model).  However, when $q>1$, the median of the distribution is  greater than the mean. As it can be observed in Figure~\ref{crit_decay_5}, there are values of $\theta$ that are subject to the median version of the paradox, but not to the mean version. This means that the term `strong paradox' introduced in~\cite{Kooti_weird} is not applicable to this case, because the mean version provides a tighter  range  of qualities in paradox, as compared to the median  version. 

Another observable trend in Figure~\ref{crit_decay_5} is that the critical values of quality are a non-decreasing functions of $q$. This can be intuitively explained as follows. When $q$ is low, the majority of the network is constituted by low quality nodes. The majority of the neighbors of a low quality node will also have low quality.  So the node does not experience the paradox with high probability. When $q$ increases, the number of nodes with higher quality increases, and a low quality node has a higher probability of being connected to those high quality nodes, which gives it a higher probability of experiencing paradox. Comparing Figure~\ref{crit_decay_8} with Figure~\ref{crit_decay_5}, we observe that as $\beta$ varies $\qtil^u$ and $\qmed^u$ do not change,  while the critical values of the QPA model get closer to those of the uncorrelated case. 
These figures only depict the results for two  values of $\beta$, due to space limitations. The trend holds for the omitted figures.   We conclude that \emph{ as $\beta$ gets larger, the correlation of the quality of a node  with  the quality of its neighbors diminishes.}

In Figure~\ref{crit_decay_21}, the critical degrees  (as defined in~\eqref{measures} and~\eqref{measures_u}) are depicted. It can be observed that as $q$ increases, $\ktil$ decreases.  Comparing Figures~\ref{crit_decay_21} and~\ref{crit_decay_24}, we observe that all the critical degrees are greater in the case of $\beta=8$ than $\beta=2$. Also the range  of the degrees who experience paradox (of any type) is wider when $\beta=8$. In both figures, we observe that the mean FP is more sensitive to changes in the quality distribution than the median FP.
\begin{figure}[!t]
        \centering
      \begin{subfigure}[b]{.5 \textwidth}
              \includegraphics[width=  \textwidth,height=4.2cm]{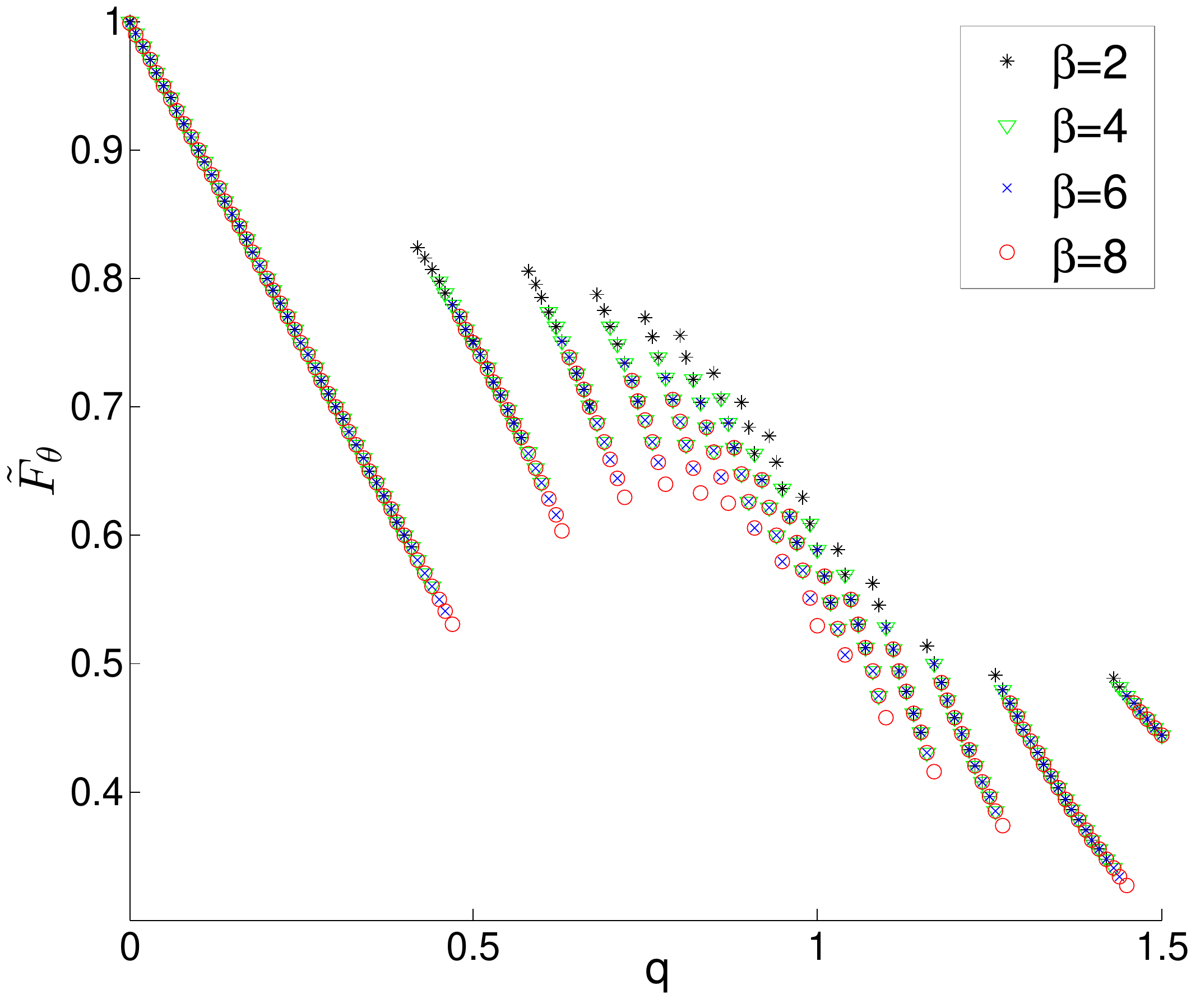}
                \caption{
mean quality paradox, $\theta_{max}=16$
}
                \label{pop_decay_1}
        \end{subfigure}%
        \begin{subfigure}[b]{0.5\textwidth}
                \includegraphics[width=  \textwidth,height=4.2cm]{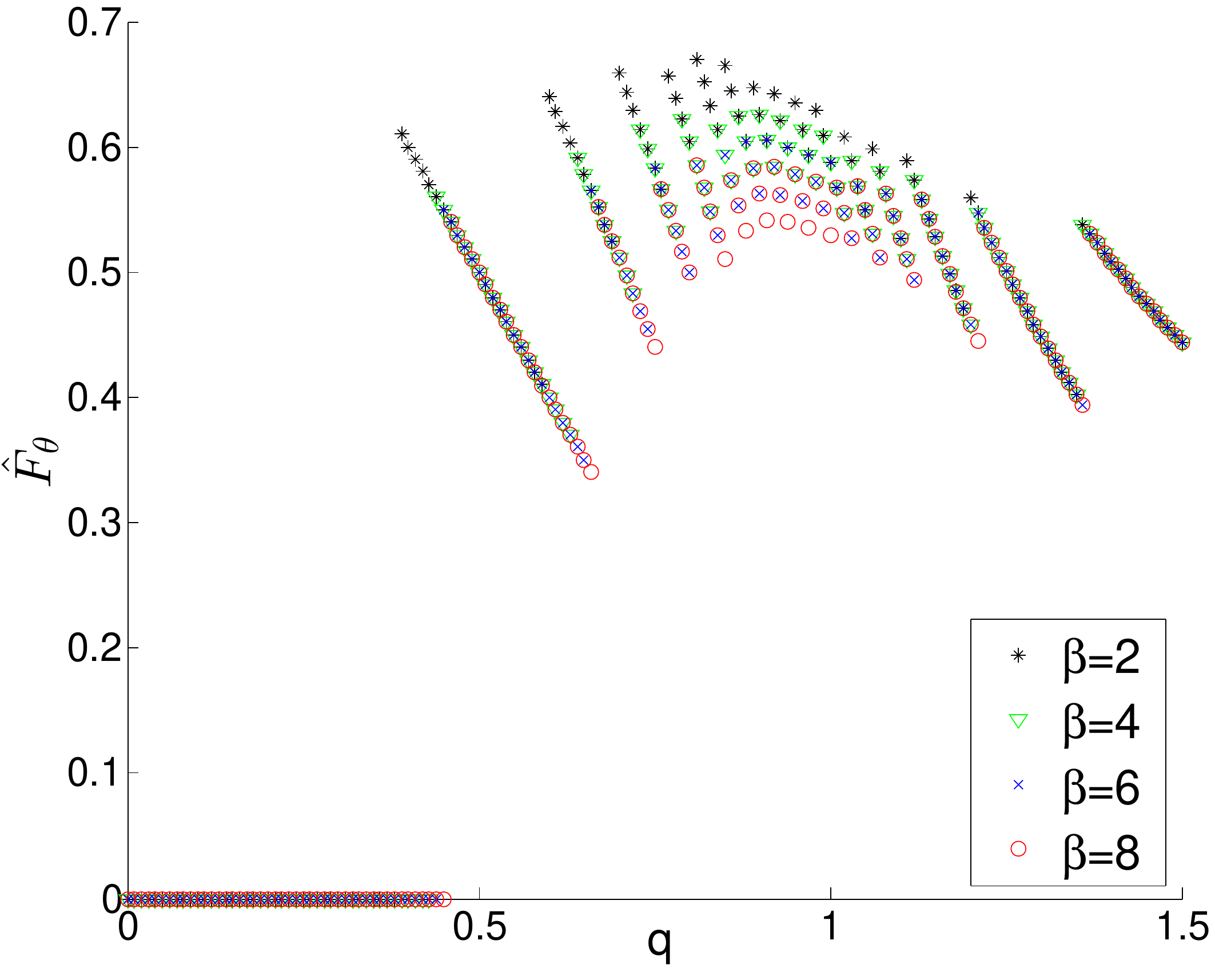}
                \caption{median quality paradox, $\theta_{max}=16$}
                \label{pop_decay_3}
        \end{subfigure}%
\\
  \begin{subfigure}[b]{.5 \textwidth}
              \includegraphics[width= \textwidth,height=4.2cm]{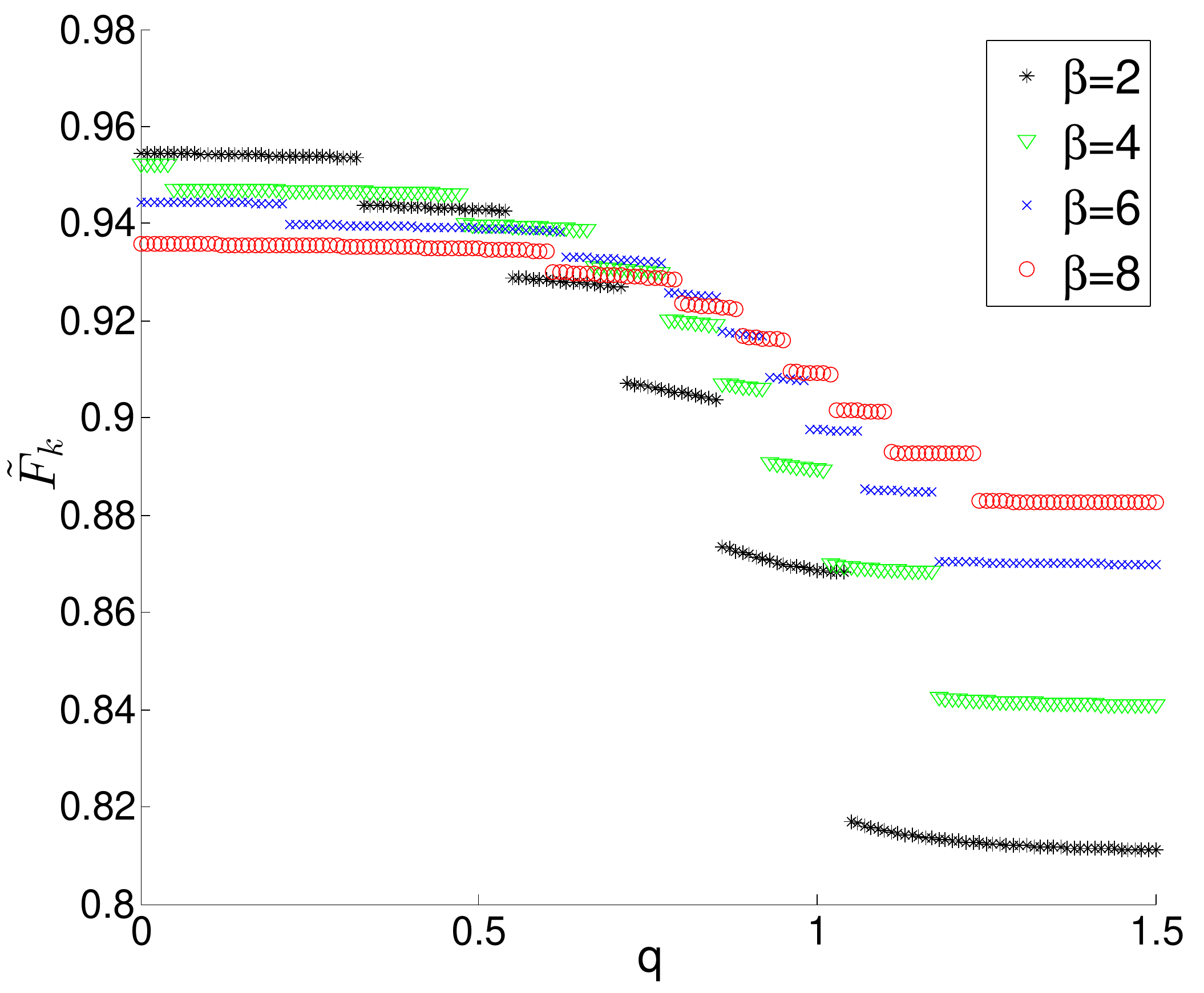}
                \caption{
mean friendship paradox, $\theta_{max}=16$
}
                \label{pop_decay_5}
        \end{subfigure}%
        \begin{subfigure}[b]{0.5\textwidth}
                \includegraphics[width=  \textwidth,height=4.2cm]{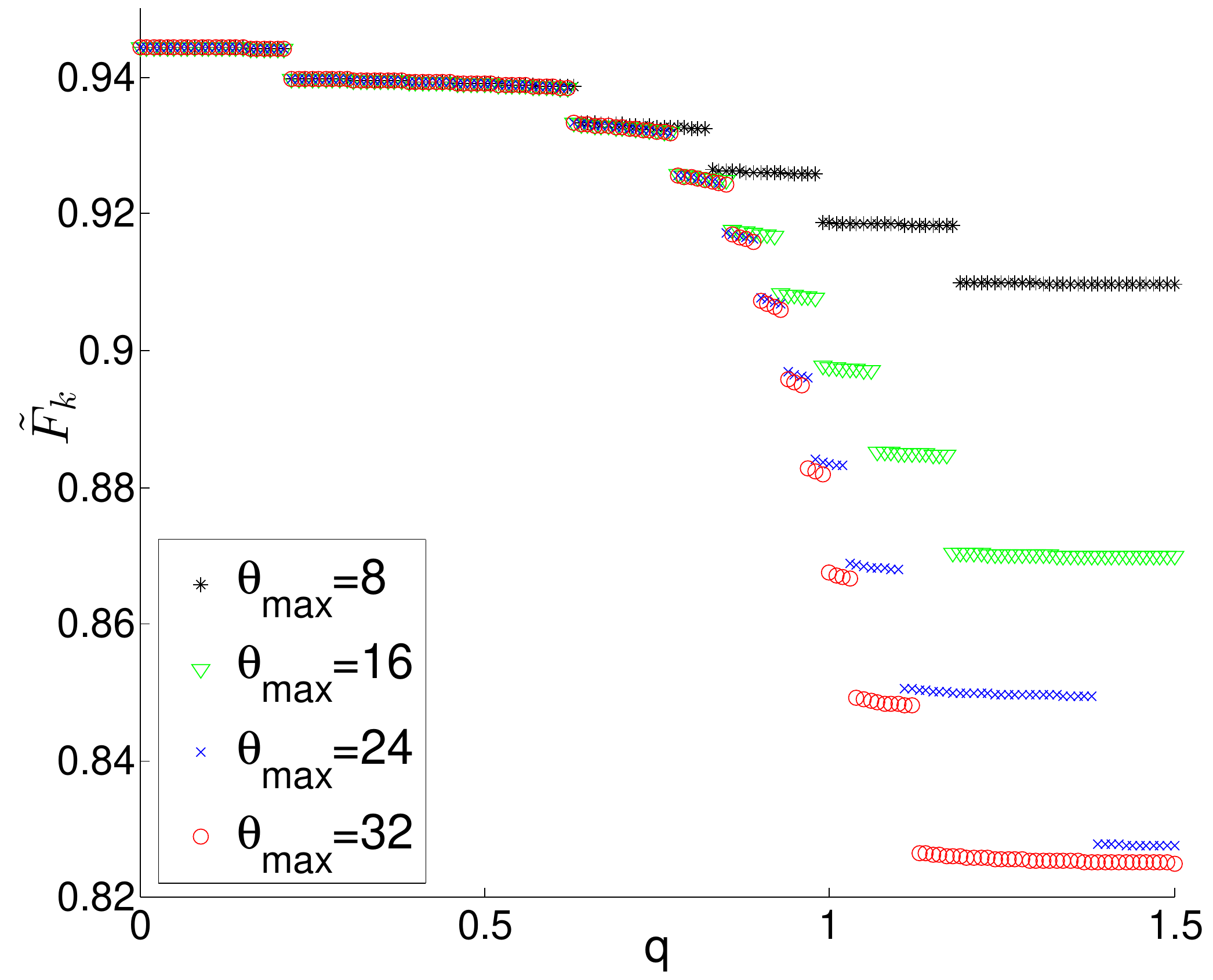}
                \caption{mean friendship paradox, $\beta=6$}
                \label{pop_decay_6}
        \end{subfigure}%
\\
 \begin{subfigure}[b]{0.5\textwidth}
                \includegraphics[width=   \textwidth,height=4.2cm]{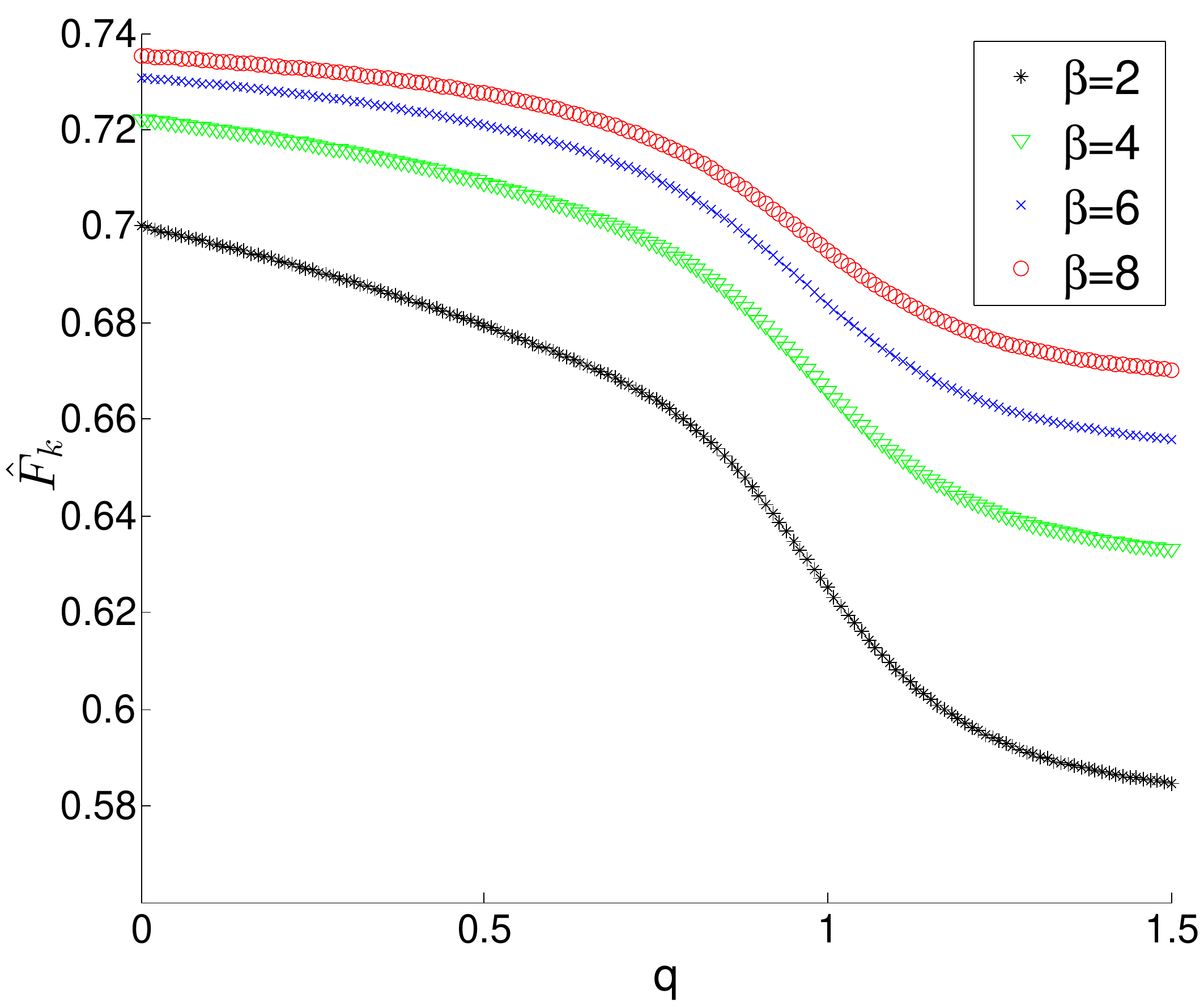}
                \caption{median friendship paradox, $\theta_{max}=16$}
                \label{pop_decay_7}
        \end{subfigure}%
        \begin{subfigure}[b]{0.5\textwidth}
                \includegraphics[width=   \textwidth,height=4.2cm]{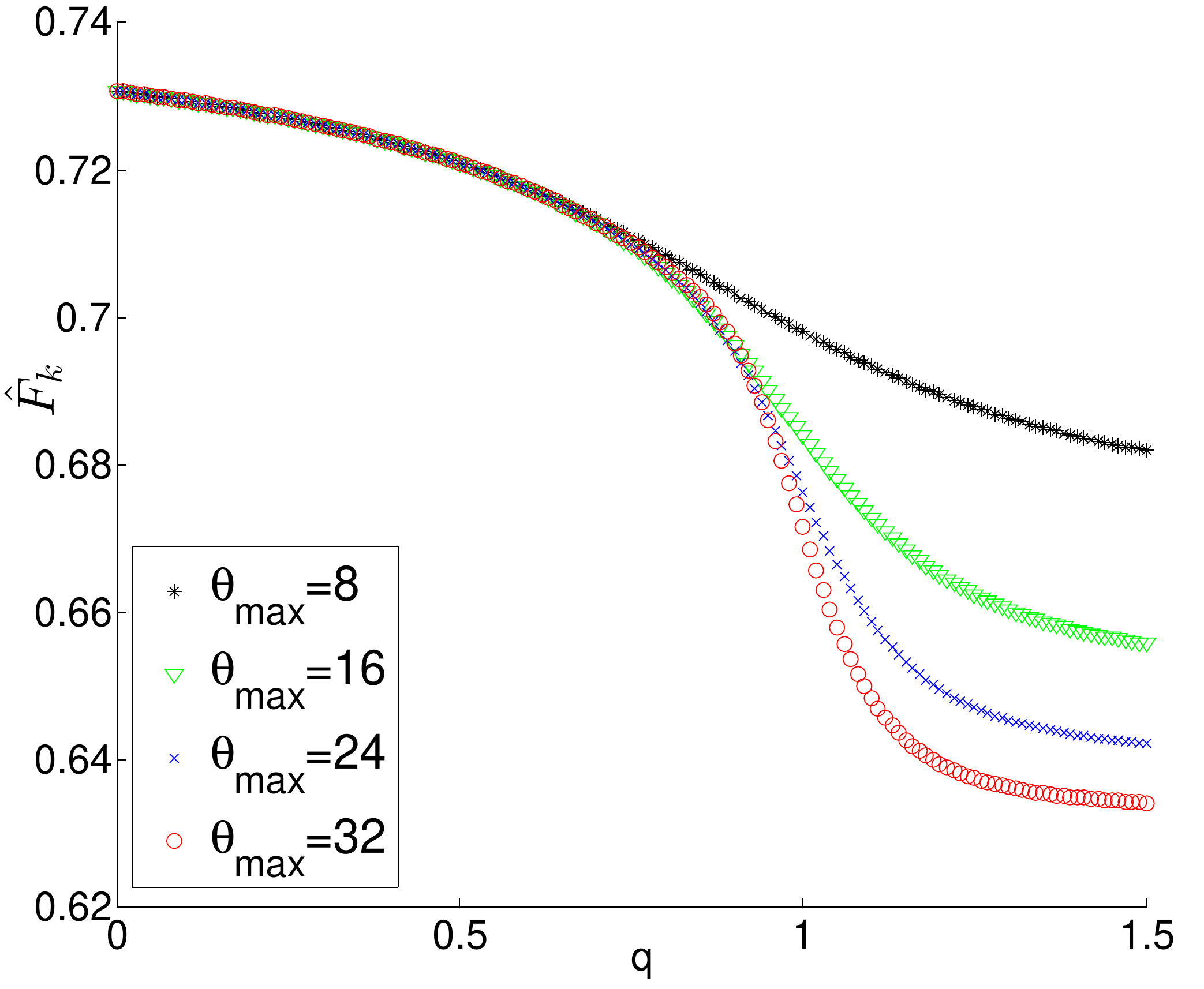}
                \caption{median friendship paradox, $\beta=6$}
                \label{pop_decay_8}
        \end{subfigure}
         \caption{
         The fraction of nodes in the quality and friendship paradoxes when the quality distribution $\rho(\theta)$ is exponential.
          }
\label{pop_decay}
\end{figure}

Figure \ref{pop_decay} depicts the fraction of nodes in the quality and friendship paradoxes (as defined in \eqref{Fs_def}) when quality distribution is exponential. From Figure~\ref{pop_decay_1} we observe that, as  $q$  increases in the vicinity of zero, $\tilde{F}_{\theta}$, the fraction of nodes experiencing the mean QP (with qualities lower than $\qtil$) decreases, because increasing $q$ increases the fraction of nodes with high qualities. The fraction $\tilde{F}_{\theta}$ has discontinuities at the values of $q$ at which $\widetilde{\theta}_c$ is incremented by one. So all the nodes  whose qualities  where equal to the new  $\qtil$ are taken into account as those who experience the mean QP, hence the abrupt jump. 

The fraction of nodes in the median QP is depicted in Figure~\ref{pop_decay_3}. It can be seen that $\hat{F}_{\theta}$ has a similar behavior to that of $\tilde{F}_{\theta}$. Each discontinuity pertains to a value of $q$ at which $\qmed$ increments. The main difference between Figures~\ref{pop_decay_1} and~\ref{pop_decay_3} is the behavior near $q=0$. In the mean QP, when almost all nodes have quality zero, even one non-zero quality neighbor  elevates the average above zero, so all those zero-quality nodes experience the mean QP. However, in the median version, at least half of the friends of a zero-quality node must have non-zero quality. Also observe that for $q<1$, we have $\widetilde{F}_{\theta} \geq \hat{F}_{\theta}$, i.e, the fraction of nodes in the mean QP is higher than the fraction of nodes in the median QP. But, for $q>1$ the inequality changes sides. 
\begin{figure}[t]
        \centering
      \begin{subfigure}[b]{.5 \textwidth}
              \includegraphics[width=.85  \textwidth,height=4.2cm]{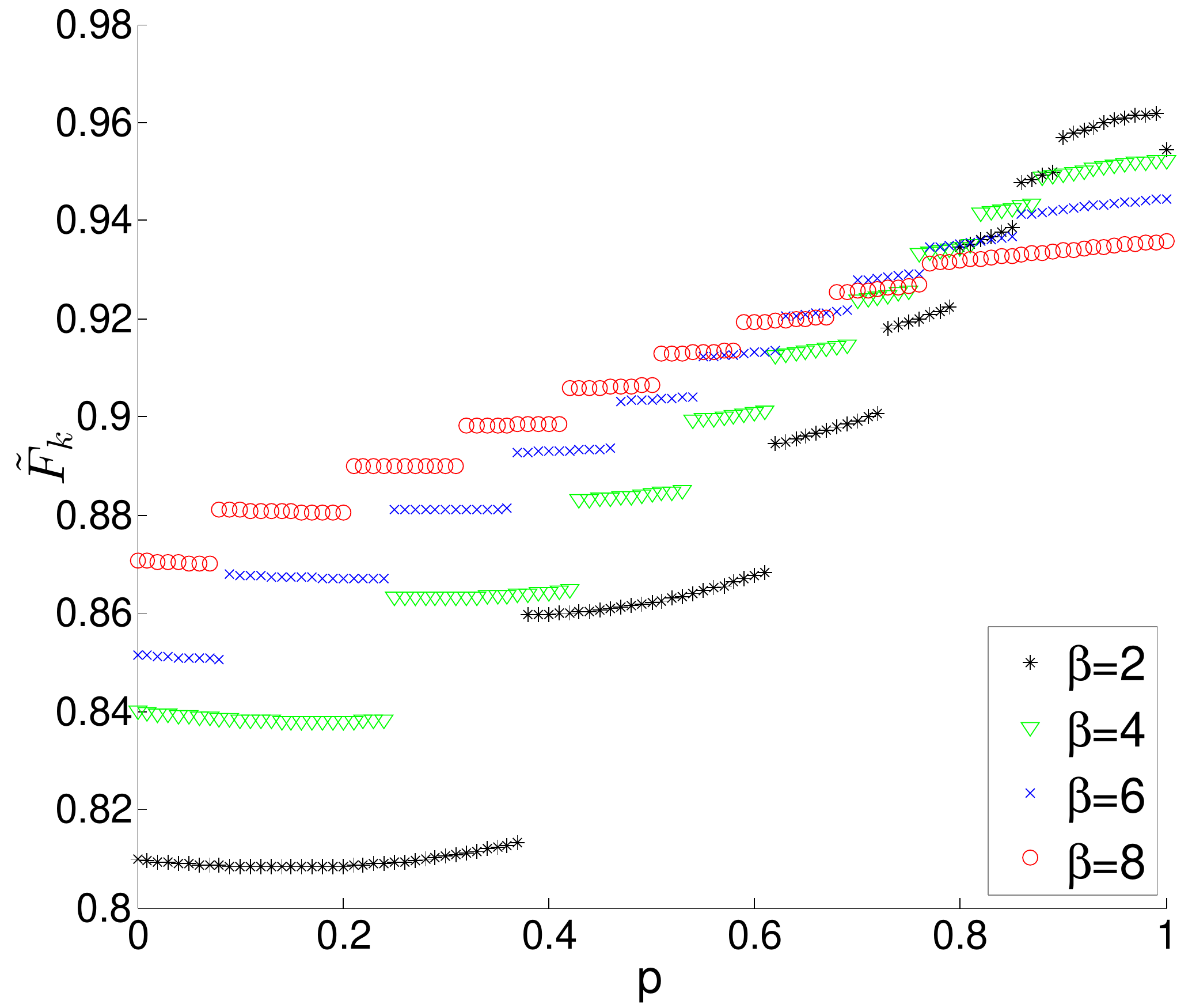}
                \caption{
mean friendship paradox, $\theta_{max}=16$
}
                \label{pop_ext_5}
        \end{subfigure}%
        \begin{subfigure}[b]{0.5\textwidth}
                \includegraphics[width=.85  \textwidth,height=4.2cm]{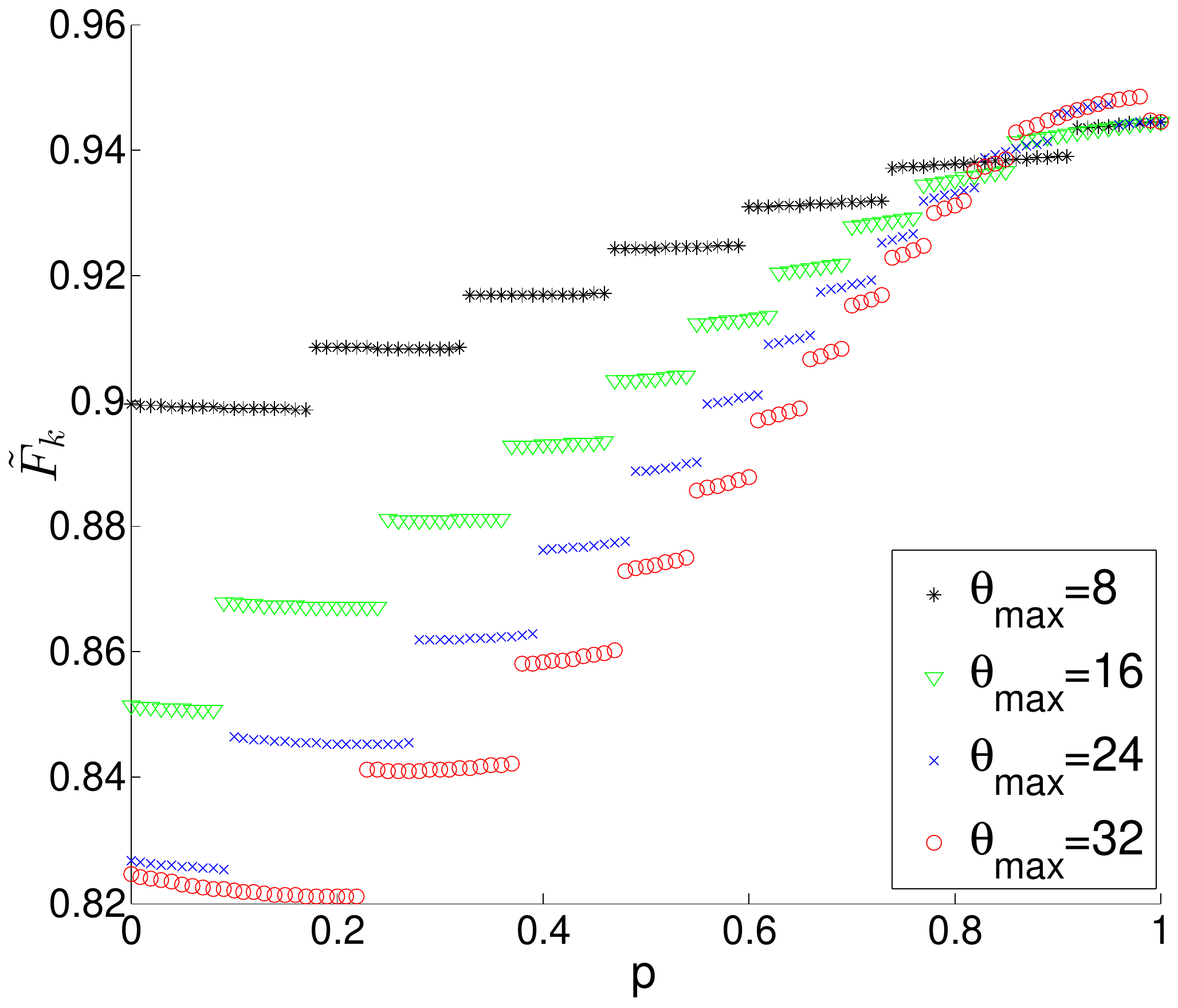}
                \caption{mean friendship paradox, $\beta=6$}
                \label{pop_ext_6}
        \end{subfigure}%
\\
 \begin{subfigure}[b]{0.5\textwidth}
                \includegraphics[width= .85  \textwidth,height=4.2cm]{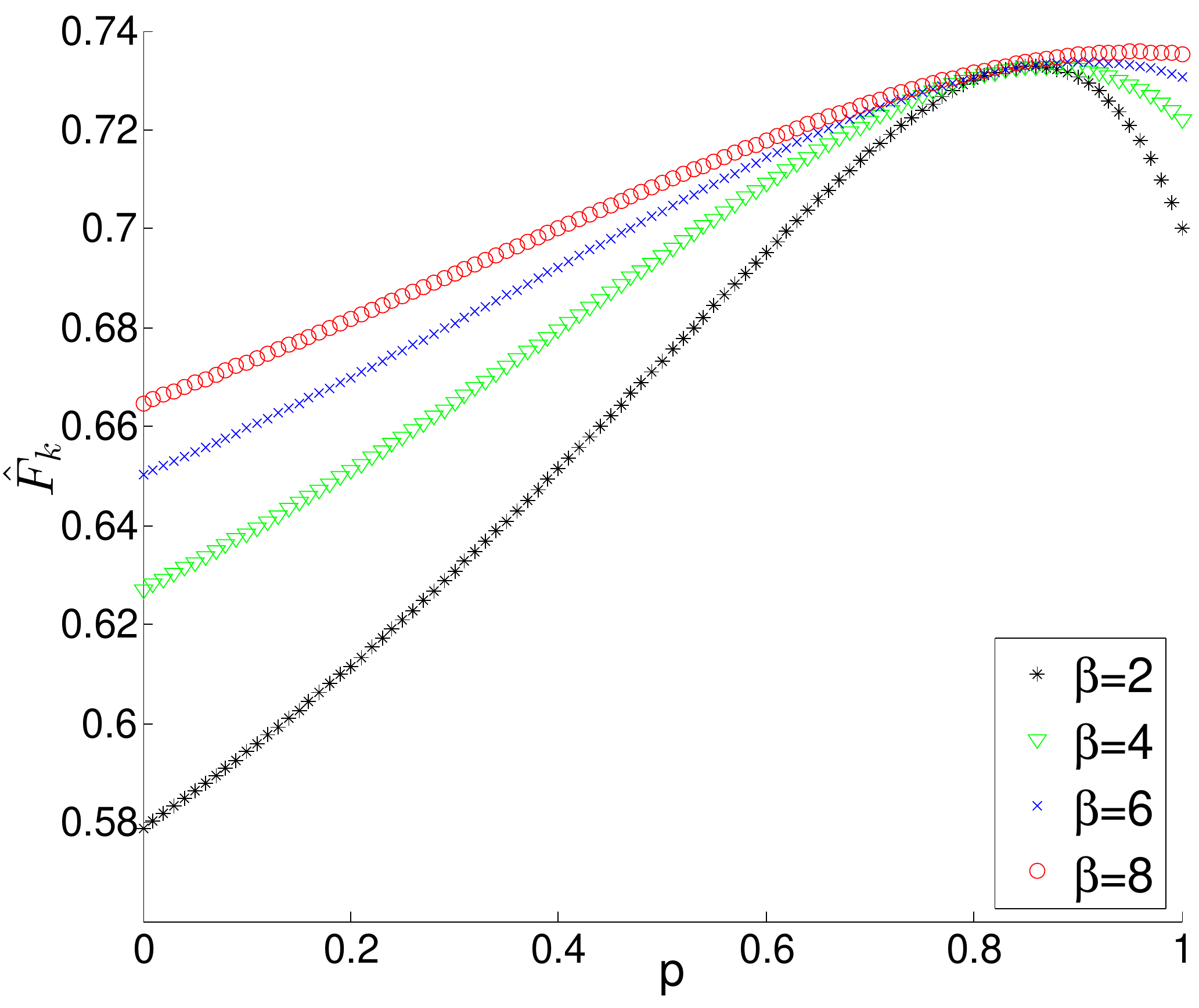}
                \caption{median friendship paradox, $\theta_{max}=16$}
                \label{pop_ext_7}
        \end{subfigure}%
        \begin{subfigure}[b]{0.5\textwidth}
                \includegraphics[width= .85  \textwidth,height=4.2cm]{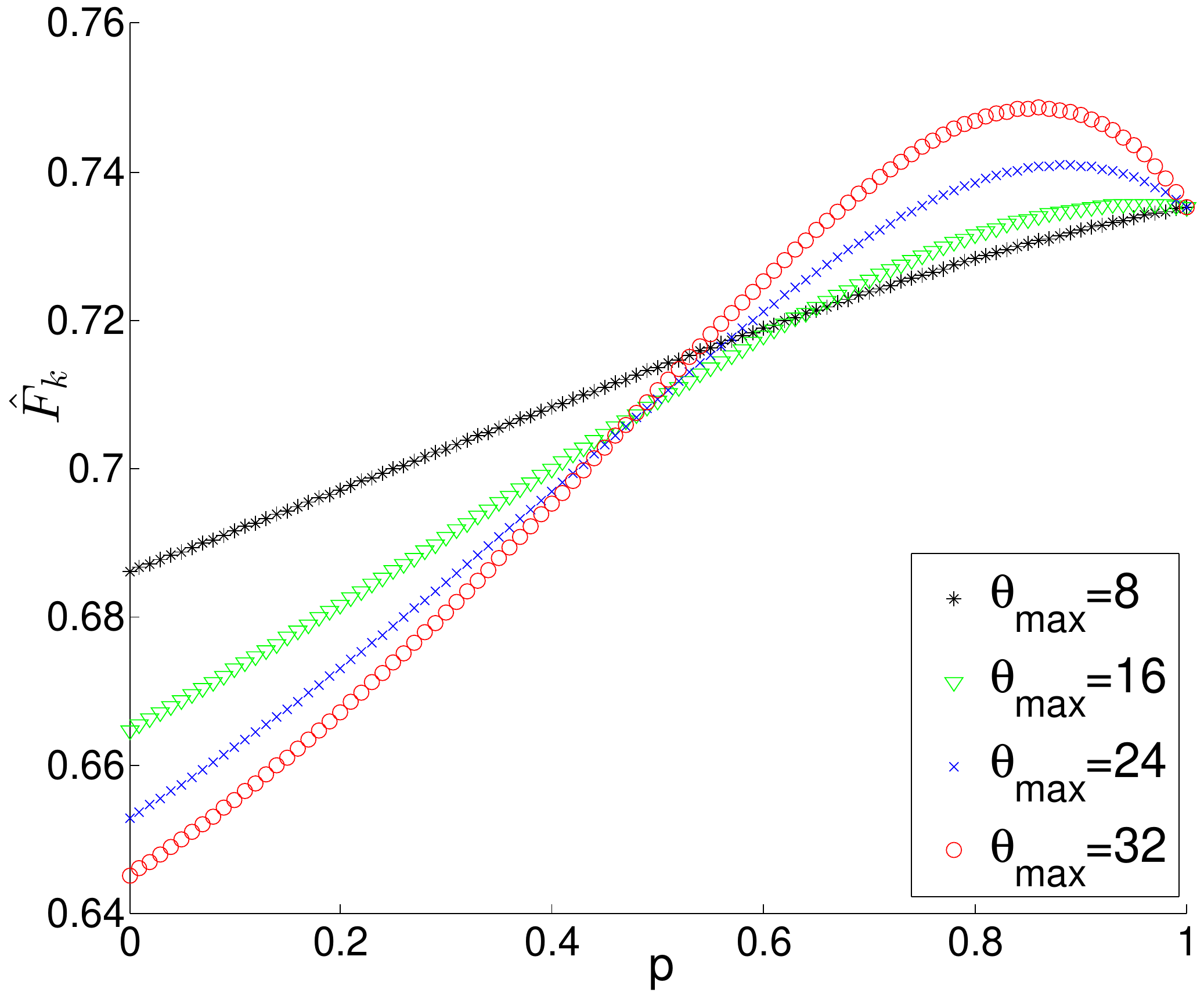}
                \caption{median friendship paradox, $\beta=8$}
                \label{pop_ext_8}
        \end{subfigure}
         \caption{
         The fraction of nodes in the friendship paradox when the node quality distribution $\rho(\theta)$ is Bernoulli.
          }
\label{pop_ext}
\end{figure}

In Figures~\ref{pop_decay_5} and~\ref{pop_decay_6}, it can be observed that  \emph{for all values of $\beta$ and $\thmax$, the majority of the nodes (over $80\%$) experience the mean FP. }  Also, as $q$ increases, $\widetilde{F}_k$ decreases. It means that the quality distribution affects the FP that depends solely on degrees. Through the quality-dependant network growth mechanism, the degree distribution, and hence the conditions under which a node experiences the FP, depend on the quality distribution. Also, it is observed in Figure~\ref{pop_decay_5} that as $\beta$ increases, the  sensitivity  of $\widetilde{F}_k$ to variations of $q$ decreases. This means that as the initial degree of nodes increases, the effect of the quality distribution on the FP diminishes. Because as $\beta$ increases the final degrees of nodes increase, and for larger degrees  ${k+\theta}$ is dominated by $k$; varying $\theta$ has less of an effect. Conversely, in Figure~\ref{pop_decay_6}, as $\thmax$ increases, the  sensitivity  of $\widetilde{F}_k$ to variations of $q$ increases. As the  range of possible qualities becomes wider, the probability of having high values of $\theta$ that have significant roles in  ${k+\theta}$  increases. 

 In Figures~\ref{pop_decay_7} and~\ref{pop_decay_8}, we observe that as $q$ increases, $\hat{F}_k$ (the fraction of nodes experiencing the median FP) decreases. This is similar to the trend observed for $\widetilde{F}_k$ in Figures~\ref{pop_decay_5} and~\ref{pop_decay_6}. From Figure~\ref{pop_decay_7} we observe that $\hat{F}_k$ increases as $\beta$ increases. From Figure~\ref{pop_decay_8} we observe that  for a range of decay factors (up to around $q=0.7$), $\thmax$ does not have a significant effect on  $\hat{F}_k$, but beyond that point, $\hat{F}_k$ decreases as $\theta_{\max}$ increases.  Also, comparing Figures~\ref{pop_decay_7} and~\ref{pop_decay_8} with Figures~\ref{pop_decay_5} and~\ref{pop_decay_6}, we assert that $\hat{F}_k \leq \widetilde{F}_k$. In other words,\emph{  the  median  FP is always stronger than the  mean  FP, regardless of the quality distribution}. 

The fraction of nodes experiencing the FP when the quality distribution is Bernoulli are depicted in Figure~\ref{pop_ext}. From Figures~\ref{pop_ext_5} and~\ref{pop_ext_6} we observe that as $p$ increases, $\widetilde{F}_k$  (the fraction of nodes experiencing the mean FP) increases. From Figure~\ref{pop_ext_5} we deduce that as $\beta$ increases, the sensitivity of $\widetilde{F}_k$ to variations of $p$  decreases. Also, in Figure~\ref{pop_ext_6} it is observed that as $\thmax$  increases, the sensitivity of $\widetilde{F}_k$ to variations of $p$   increases (similar to Figures~\ref{pop_decay_5} and~\ref{pop_decay_6}).

From Figure~\ref{pop_ext_7} we observe that as $\beta$ increases, $\hat{F}_k$ (the fraction of nodes experiencing the median FP) increases. From Figure~\ref{pop_ext_8} we observe that as $\thmax$ increases, the sensitivity of $\hat{F}_k$ to the variations of $p$ increases. Comparing Figures~\ref{pop_ext_5} and~\ref{pop_ext_6} with Figures~\ref{pop_ext_7} and~\ref{pop_ext_8} we deduce that for each value of $p$, we have ${\hat{F}_k \leq \widetilde{F}_k}$, i.e., \emph{the fraction of nodes experiencing the mean FP is higher than nodes in the median FP regardless of the quality distribution}.
\vspace{-2.5mm}
\section{Summary and Future Work}
\vspace{-2.5mm}
In this paper we studied the friendship and the generalized friendship paradoxes on networks grown under a quality-based preferential attachment scheme. To this end, we introduced measures, such as quality and degree critical values, and fraction of nodes that experience each paradox. In each case, we considered  the mean and the median to characterize the paradox. We compared the results to the uncorrelated network where the qualities and degrees of neighbors are uncorrelated.  We considered  Bernoulli and exponential distributions for qualities.

For the exponential quality distribution, the critical quality of the uncorrelated case is always smaller than that of the QPA model. This means that the range of possible values of the quality that experience paradox is wider in the QPA model than in the uncorrelated case. We  also observed that as $\beta$ increases, the nearest-neighbor quality correlation decreases. In other words, the critical values of the proposed model converge to those  of the uncorrelated case. For the exponential quality distribution we also observe that when $q<1$ (which makes the median smaller than the mean), the median QP is stronger than the mean QP for both the QPA model and the uncorrelated case. The converse is true for $q>1$. For all values of $\beta,\thmax$, over $80\%$ of nodes experience the mean FP. We observed that changing the distribution of qualities affects the FP (in addition to the QP). This effect is strengthened when $\beta$ decreases or when $\thmax$ increases. Also, it was observed that regardless of the quality distribution, the median FP is always stronger than the mean FP. 

Plausible extensions of the present contribution are as follows. We can apply the measures introduced here to real networks, and compare the results, and also compare them  with  networks  synthesized with  arbitrary quality distributions. This enables us to investigate what type of quality distribution best characterizes a given network.

\end{document}